\documentclass[twocolumn,showpacs,aps,prl,superscriptaddress,floatfix]{revtex4}

\usepackage{graphicx}
\usepackage{dcolumn}
\usepackage{amsmath}
\usepackage{epsfig}

\RequirePackage{xspace}

\begin{document}

\title{B-Reconstruction Methods via Geometro-Kinematic Constraints (I)}

\author{A. Dima}
\affiliation{Department of Engineering, University of Liverpool, Brownlow Hill 
L69 3GH Liverpool, UK}
\author{M. Dima}
\affiliation{Institute
for Physics and Nuclear Engineering, Str. Atomistilor 407, P.O. Box MG-6, R-76900 Bucharest, Romania
}

\date{\today}       

\begin{abstract}
Decay channels with attractive branching ratios, or
interesting physics, are recovered by substituting
``missing" particles ($\gamma$, $\nu$, $\pi^0$, etc)
 with combined geometric and kinematic constraints.
The ``Sliding Vertex" method
is shown in this part-I, for reconstructing strongly 
boosted 
 B$^0_s$
decays - at the LHC.
\end{abstract}


\maketitle

Precision B-Physics in today's experiments~\cite{bexp}
relies heavily on exclusively reconstructed decay 
chains.
There are however decay modes
that are not readily exclusive-reconstructable,
but very attractive from the view point of the
physics, or that of a large 
branching ratio - impaired
in reconstruction usually by
 missing particles.
Methods for constraining and recovering information
in missing particle events
have been explored before
with kinematic fitting~\cite{bib:avery,bib:kinf}
and in a variety of other contexts, a
good collection thereof being~\cite{bib:k1,bib:k2}.

Consider for instance the B$^0_s$ $\rightarrow$ D$^-_s$ $K^+$
decay with the subsequent 
D$^-_s$ $\rightarrow$ K$^+$ K$^-$ $\pi^-$ ($\pi^0$)
decay.
This mode is simple and attractive for
measuring $\gamma_{_{\rm CKM}}$, however 
its branching ratio is rather fair.
If we consider the related mode, with $\pi^0$
in the final state, the branching ratio is ca. 3.5 times
larger, a real feat.
Evidently, the $\pi^0$ can be reconstructed in the E-calorimeter,
albeit with less resolution than for tracking.
In principle the reconstruction with the
$\pi^0$ is very attractive,
but seemingly somewhat impractical.

There is however enough information in the detector to
reconstruct the decay without the $\pi^0$: first there
are the 4-momentum conservation laws in the B$_{\rm VTX}$
and D$_{\rm VTX}$, and second, the B$_{\rm VTX}$ must
lie on the $\pi^+$ track.
\begin{figure}[b]
\centering
\includegraphics[width=8.0cm]{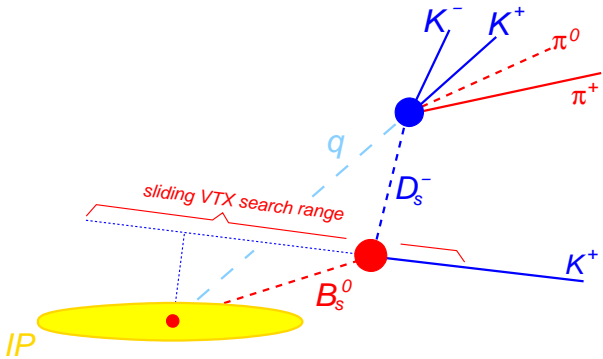}
\caption {Topology of B$^0_s$ $\rightarrow$ D$^-_s$ $K^+$
decay with the subsequent
D$^-_s$ $\rightarrow$ K$^+$ K$^-$ $\pi^-$ ($\pi^0$).
The tracks are shown
in the absence of magnetic field - such as for LHCb,
although, given the small track bending over the
vertexing region, a solution
in the form of an ${\cal O}_{III}$ helical correction
exists
also for ATLAS.
}
\label{fig:b}
\end{figure}
Figure \ref{fig:b} shows the topology of the event
(in the absence of magnetic field - such as for LHCb,
although, given the small track bending over the
vertexing region, a solution
in the form of an ${\cal O}_{III}$ helical correction
exists
also for ATLAS).
The evident question arises, whether
such methods reject not only background,
but also ``sister physics" (in the case of the
above mode 
the significantly 
more abundant B$^0_s$ $\rightarrow$ D$^-_s$ $\pi^+$
decay, with the $\pi^+$ mistaken by the PID
system as a K$^+$), which play a more dominant role.
Such is the case of decays coming from
B$^0_d$ 
which to a large degree shadow (ca. 3:1) the B$^0_s$ decays
through similar mass ratios
between mother and daughters.
From the start it should be stated that
this is the most difficult part and 
that possible solutions lie in the
avenues of tracking/vertexing
resolutions, of (semi)-leading (bi)-particle
effects in the aforementioned decays and
of different branching ratios (sometimes smaller) for B$^0_d$
for the exact event topology as B$^0_s$.
In fact B$^0_d$ contaminates B$^0_s$ events more heavily
with different topologies (where 1 track is lost, or
which have 1-2 extra $\pi^0$'s).
 
{\bf Kinematic Constraints} - are the 4-momentum conservation
laws in the B$_{\rm VTX}$ and D$_{\rm VTX}$. For the B$_{\rm VTX}$
the useful part is:
\begin{eqnarray}
 (E_\pi+E_D)^2 &=& M_B^2 + (p_D^2+p_\pi^2+2p_Dp_\pi cos\theta_{D\pi}) \cr \cr
{\rm where \, \, \, \, } cos\theta_{D\pi} &=& \frac{1}{p_D}(E_D/\beta_\pi-Q_{D\pi}) \cr \cr
{\rm with \, \, \, \, } Q_{D\pi} &=& \frac{1}{p_\pi}\Delta_{D\pi}^2 \cr \cr
{\rm and \, \, \, \, \, \, } \Delta_{D\pi}^2 &=& 
\frac{1}{2}(M_B^2-M_D^2-m_\pi^2)
\end{eqnarray}
and for the D$_{\rm VTX}$:
\begin{eqnarray}
 (E_\pi-E_v)^2 &=& m_0^2 + (p_D^2+p_v^2-2p_Dp_v cos\theta_{Dv}) \cr \cr
{\rm where \, \, \, \, } cos\theta_{Dv} &=& \frac{1}{p_D}(E_D/\beta_v-Q_{Dv}) \cr \cr
{\rm with \, \, \, \, } Q_{Dv} &=& \frac{1}{p_v}\Delta_{Dv}^2 \cr \cr
{\rm and \, \, \, \, \, \, } \Delta_{Dv}^2 &=&
\frac{1}{2}(M_D^2+m_v^2-m_0^2)
\end{eqnarray}
where ``0" is the missing neutral, ``v" the sum of visible
particles in the D$_{\rm VTX}$ and the rest of notations are self-evident.

{\bf Geometric Constraints} - are supplying the missing, 3$^{\rm rd}$
equation to the above set. Due to momentum conservation in the  B$_{\rm VTX}$
``D", ``$\pi$" and ``q" (see figure 1) lie in the same plane: $\vec{n}_D = 
\lambda \vec{n}_\pi + \mu \vec{n}_q$, where $\vec{n}$ are unit vectors
and $\lambda$, $\mu$ constants.
The cosines from the kinematic relations are: $cos\theta_{D\pi} =
\vec{n}_D\cdot \vec{n}\pi$ and $cos\theta_{Dv} =
\vec{n}_D\cdot \vec{n}v$, respectively:
\begin{eqnarray}
 \lambda &=& \frac{1}{\Delta}\big[(vq) cos\theta_{D\pi} - (q\pi)cos\theta_{Dv}\big] \cr \cr
\mu &=& \frac{1}{\Delta}\big[-(v\pi) cos\theta_{D\pi} + cos\theta_{Dv}\big]
\end{eqnarray}
where (ab) = $\vec{n}_a \cdot \vec{n}_b$ and
$\Delta = (vq)-(v\pi)(q\pi)$.
The ``closure" equation is $|\vec{n}_D|^2 = 1 =
\lambda^2+\mu^2 + 2(q\pi)\lambda \mu$.

In terms of the geometric constraints the kinematic section condenses to
$E_D^2 - M_D^2 = (\lambda p_D)^2 + (\mu p_D)^2 +
2(q\pi)(\lambda p_D)(\mu p_D)$ which can be solved 
in favor of $E_D$ as a second order equation. The two fold
ambiguity resulting therefrom is lifted through a (2D)
pointback to IP criterion for the $B_s^0$ (after vertex
determination).

At this point all 
kinematic quantities are 
known\footnote{Historically the solution was reached 
``sliding" the position of the B$_{\rm VTX}$ along the
$\pi$ track until the $B_s^0$ mass and IP-pointback
where simultaneously met - which also lifted the
two-fold ambiguity. In the newer version of ``Sliding VTX"
which we are here presenting,
the B$_{\rm VTX}$ is determined separately, after
the kinematic quantities are known. Ancient history
has it that ``Sliding VTX" started in the equations
of the ``p$_\perp$ corrected B-mass" of SLD, SLAC-PUB-7170, hence
its initial ``SLD-ing" approach.}.

{\bf Vertex Determination} - once $E_D$ is known the 3D
vectors of all particles are known. The B$_{\rm VTX}$
$\vec{b}$ is the locus that takes the best-shot at the:
\begin{enumerate}
\item{IP in the  $\vec{n}_B$ direction:
 $\vec{b}-\lambda_B\vec{n}_B = IP \pm \sigma_{IP}$}
\item{D$_{\rm VTX}$ in the  $\vec{n}_D$ direction:
 $\vec{b}-\lambda_D\vec{n}_D = D_{\rm VTX} \pm \sigma_{D}$}
\item{$\pi^+$ track 1$\sigma_\pi$ tube:
 $\vec{b}-\lambda_\pi\vec{n}_\pi = \vec{r}_{0\pi} \pm \sigma_{D}$}
\end{enumerate}
Mathematically this means - using the ${\boldmath \sigma}_i$
error matrices:
\begin{eqnarray}
&\langle& \vec{b} - \vec{v}_B - \lambda_B \vec{n}_B \, \, | 
\, \, {\boldmath \sigma}_{IP}^{-2} \, \, | \, \, 
\vec{b} - \vec{v}_B - \lambda_B \vec{n}_B \, \, \rangle + \cr
&\langle& \vec{b} - \vec{v}_D - \lambda_D \vec{n}_D \, \, \! |
\, \, {\boldmath \sigma}_{D}^{-2} \, \, | \, \, 
\vec{b} - \vec{v}_D - \lambda_D \vec{n}_D \, \, \rangle + \cr
&\langle& \vec{b} - \vec{r}_{0\pi} - \lambda_\pi \vec{n}_\pi \, \, |
\, \, {\boldmath \sigma}_{\pi}^{-2} \, \, |\, \, 
\vec{b} - \vec{r}_{0\pi} - \lambda_\pi \vec{n}_\pi \, \, \rangle = min
\end{eqnarray}
By differentiating to find the minimum, the ``sliding"
along each direction is:
\begin{equation}
\lambda_i = \langle \,\, \vec{n}_i \, \, |
\, \, {\boldmath \sigma}_i^{-2} \, \, |
\, \, \vec{b}-\vec{v}_i \, \, \rangle \, \, /
\, \, \langle \, \, \vec{n}_i \, \, |
\, \, {\boldmath \sigma}_i^{-2} \, \, |
\vec{n}_i \, \, \rangle
\end{equation}
The vertex solution is then:
\begin{equation}
|\, \, \vec{b} \, \, \rangle =
\bigg[\sum_i
{\boldmath \sigma}_i^{-2}
\bigg({\boldmath 1} - 
\frac{|\, \, \vec{n}_i \, \, \rangle
\langle \, \, \vec{n}_i \, \, |
{\boldmath \sigma}_i^{-2}}
{\langle \, \, \vec{n}_i \, \, |
\, \, {\boldmath \sigma}_i^{-2} \, \, |
\vec{n}_i \, \, \rangle
}\bigg) \bigg]^{-1}
\, \, | \, \, \vec{w} \, \, \rangle
\end{equation}
where:
\begin{equation}
|\, \, \vec{w} \, \, \rangle =
\sum_i
{\boldmath \sigma}_i^{-2}
\bigg({\boldmath 1} -
\frac{|\, \, \vec{n}_i \, \, \rangle
\langle \, \, \vec{n}_i \, \, |
{\boldmath \sigma}_i^{-2}}
{\langle \, \, \vec{n}_i \, \, |
\, \, {\boldmath \sigma}_i^{-2} \, \, |
\vec{n}_i \, \, \rangle
}\bigg) 
\, \, | \, \, \vec{v}_i \, \, \rangle
\end{equation}

The math in itself is straightforward, however
computer implementation proved to be somewhat of
a hassle. Code that implements vectors and matrices
is rather slow due to multiple inheritances, unnecessary
functions, etc. We coded our own MXV4~\cite{bib:mxv4}
namespace that addressed these issues 
and performed in speed.

One special mention is with respect to
 the $\pi$ track ${\boldmath \sigma}_\pi^{-2}$ matrix.
This is not immediate two-fold: writing the
${\boldmath \sigma}_\pi$ and secondly, the tube is
not cylindrical, rather elliptical in cross-section.
This is due to to the fact that the track
is determined by (approximately) circular errors
in the VELO planes of LHCb (prototype addresse) which stand
vertically.
(This of course remains to be solved exactly for each
detector where it is applied.)
The $\pi$ track tube is thus an infinitely long
ellipsoid with unequal cross-section major axes:
\begin{eqnarray}
{\boldmath \sigma}_\pi^{-2}  &=& 
\frac{1}{\underbrace{\sigma_{long}^2}_{=\infty}}
\, |\, \, \vec{n}_\pi \, \, \rangle
\langle \, \, \vec{n}_\pi \, \, |
\, \, +
\cr 
\cr
&\!&
{\boldmath \sigma}_{minor}^{-2}
\frac{|\, \, \vec{e}_z\times\vec{n}_\pi \, \, \rangle
\langle \, \, \vec{e}_z \times \vec{n}_\pi \, \, |}
{1- \langle \, \, \vec{e}_z \, \, | 
\, \, \vec{n}_\pi \, \, \rangle^2} \, \, +
\cr
\cr
&\!&
{\boldmath \sigma}_{major}^{-2}
\frac{|\, \, \vec{n}_\pi \times \vec{e}_z\times\vec{n}_\pi \, \, \rangle
\langle \, \, \vec{n}_\pi \times \vec{e}_z \times \vec{n}_\pi \, \, |}
{1- \langle \, \, \vec{e}_z \, \, |
\, \, \vec{n}_\pi \, \, \rangle^2}
\end{eqnarray}
where $ \sigma_{\rm minor}^{-2} = \sigma_\pi^{-2}$ and
$ \sigma_{\rm major}^{-2} = (\sigma_\pi/\langle \, \, \vec{e}_z \, \, |
\, \, \vec{n}_\pi \, \, \rangle)^{-2}$.
With this, the $\pi$ track contribution in the B$_{\rm VTX}$
calculation reduces significantly (and this needs to be so 
implemented in the code in order to avoid singularities).
\begin{figure}[htb]
\centering
\includegraphics[width=8.0cm]{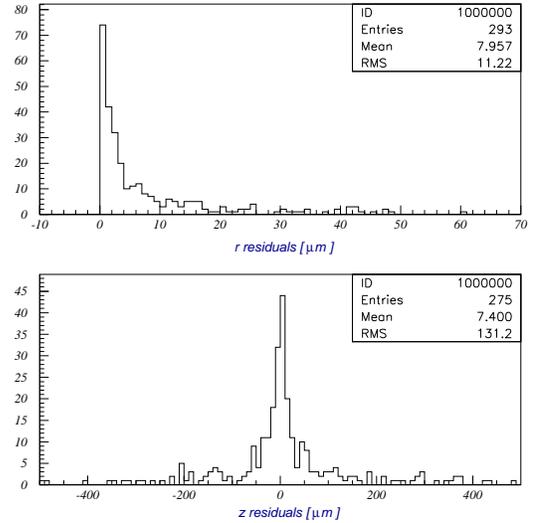}
\caption {The
xy- (top) and z-residuals (bottom) for the reconstructed B$_{\rm VTX}$
with respect to the Monte Carlo position. All dimensions are
 in $\mu$m.
}
\label{fig:bvtx}
\end{figure}

We tested the code on PYTHIA simulated data
(smeared to give invariant mass resolutions as those 
expected in the LHCb detector).
\begin{figure}[htb]
\centering
\includegraphics[width=8.0cm]{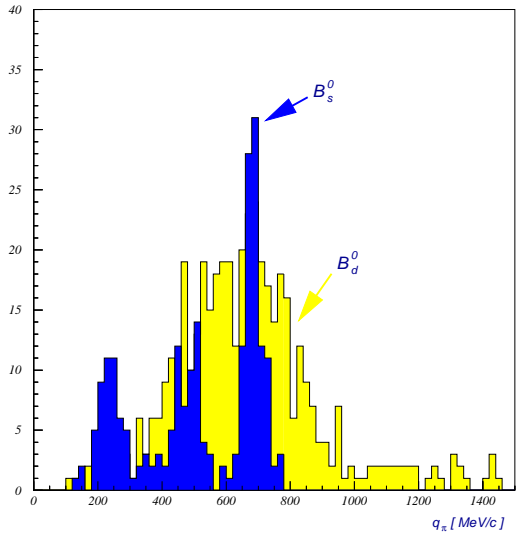}
\caption {Momentum of $\pi^-$
coming from
D$_s^-$ (blue) vs. that from D$^-$.
The spectrum is harder for the latter,
as the other two components do not
necessarily each contain a valence quark
of the decaying particle, leading to a more
equitable energy distribution among daughter
particles. This is not the case
for the two kaons comming from
a D$_s^-$ decay, which together take a greater
share of the decay energy (leading-particle effect~\cite{bib:dima}.}
\label{fig:lead}
\end{figure}

The 
xy- and z-residuals for the reconstructed B$_{\rm VTX}$
with respect to the Monte Carlo position are shown in
figure \ref{fig:bvtx}.
With an rms of 130 $\mu$m and a central part of 30 $\mu$m,
 the z-resolution is very good compared to typical
expectations on the order of 200-300 $\mu$m.
The xy$_{\rm rms}$ is 11 $\mu$m with a central core
of 4 $\mu$m, an excellent value.
As mentioned, it may be possible to reconstruct the mode,
but then there are further problems
associated with 
background suppression, in particular that of similar-physics.
In principle soft/QCD backgrounds are kinematically
very far off and eliminable. The events that are a concern
are those from B$_d^0$, insomuch not those of the same topology,
rather those of different topology and 1-2 lost tracks, and/or
mis-ID tracks.

{\bf Leading particle effects} - figure
\ref{fig:lead} 
shows the $\pi^-$ momentum (yellow) from D$^-$ $\rightarrow$
$\tilde{\pi}^-$ 
$\pi^-$
K$^+$
of the decay B$_d^0$ $\rightarrow$
D$^-$ $\pi^+$, where 
$\tilde{\pi}^-$
is mis-ID'ed as a kaon.
Charge does not help in distinguishing, as both $B$ and
$\bar{B}$
can be present, mass differences are small, hence
the only factor that can still play some role would be 
one due to dynamics.
Since the CM decay energy is 0.7-0.8 GeV some leading-particle~\cite{bib:dima}
effect is visible for the kaon-pair in a D$_s^-$ decay
(each holding a valence quark of the D$_s^-$).
In blue is the $\pi^-$ momentum coming from
D$_s^-$, evidently softer (less available energy, as most
is concentrated by the kaon pair).

{\bf Applications} - the method aims evidently at physics
analyses, however other useful events can also
be reconstructed: ones that have neutrinos
 (used in tagging), or
ones with many pions in one vertex (for instance
B$_d^0$
$\rightarrow$
D$^{*-}$ 
$\pi^+$
$\pi^-$
$\pi^+$
where the B$_{\rm VTX}$ has 4 pions).
The purity of such pion samples make them ideal candidates
for studies of $\pi$ $\rightarrow$ K mis-ID rates.
\\

One of us (A. Dima) is thankful to the University of Liverpool for
hosting 
under a Marie Curie fellowship. One of us (M. Dima)
acknowledges the CORINT-2 funding of the Romanian Agency
for Science and Technology.

\end{document}